# Performance Evaluation of HEVC over Broadband Networks

Saeed ur Rehman[1] and Gulistan Raja[2]

[1]Department of Electrical Engineering
University of Engineering and Technology
Taxila, Pakistan

[2]Department of Electrical Engineering
University of Engineering and Technology
Taxila, Pakistan

**Abstract**
High efficiency video coding is the current trend setting standard for coding and transmission of video content. This standard has brought in many improvements over its predecessor H264 standard. In this paper, performance evaluation of transmitting HEVC content over the simulated network environment for disaster and calamity stricken area has been under taken. In the simulation the transmitted video sequences are subjected to various error models in NS3 simulator. The effects of speed and number of hosts on the jitter and delay characteristics of the underlying network while transmitting flows of HEVC based content has been evaluated. The extent and effect of network errors on the quality of HEVC bit stream in terms PSNR has also been tested.  The results show that HEVC performs better for up to 0.001 percent network error, for up to 30 simultaneously transmitting nodes and for nodes travelling at speeds up to 100m/s.
***Keywords:*** *High efficiency video coding HEVC standard, NS3 simulator, network transmission of multimedia video, network performance evaluation, network simulation, information assurance in disaster struck area.*

## 1. Introduction

There is growing requirement of high definition videos for different applications in present age. Transmission of high definition video over the broadband networks requires dedicated bandwidth which is normally limited per user in case of both wired and wireless networks. These bandwidth limits are more stringent in case of wireless media like Wi-Fi and WiMAX networks where the channel further limits the link capacity. The transmitted video content has to pass through real life networks with multiple types of inherent errors. Optimum utilization of network requires efficient coding to reduce the number of bits required to transmit the video content.

Video compression methods have been developing over the past decades. H.264 standard, developed in 2003 [1], led to development and proliferation of video applications like video telephony, conferencing and video streaming. Transmission of HD video in today's network environments has been demanding better efficiency and network performance. Increasing video quality means increasing the bit rates. Increasing the network capacity to handle higher bitrates is one of the solutions, but a costly one. Although the network devices have also evolved with better backplanes and improved bandwidth/performance, this improvement falls short of bandwidth needs of HD and beyond HD quality video applications.

High efficiency video coding (HEVC) is a new technique which has been developed to improve the coding efficiency with better network transmission and reduced computational complexity by virtue of its support for parallel processing [2]. It claims compression efficiency of 50-70 percent over its predecessor H264, at an additional computational overhead of 2-10 times [3]. The enhanced features of HEVC are the key enablers for making high resolution video content viewable by all of us. All the work on HEVC standardization has been done under the umbrella of JVT-VC "Joint Collaborative Team on video coding" which is a joint team of ISO/IEC MPEG "Moving Picture Experts Group" and ITU-T VCEG "Video Coding Experts Group" [4].

H264, being a matured standard has been center of much research work including the mechanisms for transmitting H.264 video over IP networks and the performance evaluation of H.264/MPEG-4 over different networks [5], [6]. The HEVC standard which has similar architecture as H264, has been introduced in [3] where detailed overview of HEVC standard has been presented, covering the working, architecture and techniques used. HEVC has been proved as the better technique, with lower bit rates and support for higher resolution videos [7]. HEVC complexity and implementation analysis have been carried out in comparison to previous standards [8]. The mechanism for streaming the HEVC coded content has been presented in [9], wherein, a streaming framework for HEVC encoded content has been presented. The overview of HEVC high-level syntax and reference picture management has been explained in [10]. The end to end video quality prediction for HEVC video streaming over packet networks has been presented in [11].

Disaster area network model used in this work is related to different types of calamities encountered throughout the globe,

including floods, earthquakes, droughts and war zones. In such environments it is important tom establish communication and get overall damage and calamity picture of the affected area. Getting first hand video information is important for monitoring the relief efforts. The monitoring and surveillance of relief assets form a promising application for HEVC based network video. Furthermore, WiMAX technology provides a suitable solution (by illumination of the disaster struck area) for provision of communication network alongside satellite based terminals in such environments.

This work undertakes the evaluation of HEVC coded video transmission using WiMAX and other broadband media. HEVC coding efficiency is utilized for achieving better video response over the network. This involves the transmission of HEVC bit stream over a simulated network environment using open source NS-3 simulator [12] and the reference HEVC software [13]. The performance of video quality is evaluated under different network scenarios to see the effect on delay and jitter of the network. The objective is to get results that are able to reflect the behavior of HEVC coded video after being subjected to network conditions.

This paper is organized as follows. Section I is the introduction. Section II introduces the disaster area network model. Section III covers the simulation setup and the results. Section IV concludes the paper.

## 2. Overview of Disaster Area Network Model

The system architecture is shown in Figure 1, where the network access is provided to a disaster relief organization in a disaster/calamity stricken area. The organization consists of a central relief office from where all the relief activity is controlled and field teams consisting of relief workers and their vehicles. At the central relief office and its vicinity the employees/relief workers are provided network access through WLAN (IEEE 802.11, Wi-Fi) and LAN (IEEE 802.3, Ethernet) based devices. The field teams are provided with network access through WiMAX (IEEE 802.16) based user terminals. The control office is connected via 100Mbps backhaul Ethernet link to the WiMAX BTS tower installed in the vicinity of calamity area.

The network users are the relief and emergency teams equipped with handheld and vehicle mounted devices. These teams are sending and exchanging video information with the control center. This information includes the captured and received video streams. The received video information may contain the progress of other teams and video from the adjacent or priority area. This may include telemedicine video streams for the field medical workers and weather, geographic or any other important information concerning the relief operation. At the headquarters this information may be valuable for the optimal utilization and tasking of the relief teams depending upon the damage in a particular area.

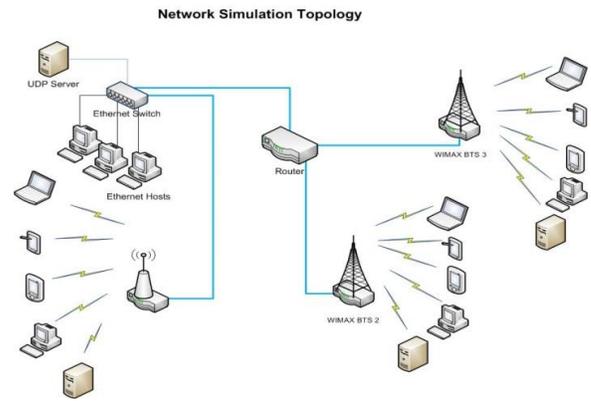

Fig. 1 The Simulated Network Topology.

The network model considered for the analysis includes up to 30 WiMAX mobile nodes. These nodes are simulated in an area of 4500x4500 m. These mobile nodes connect with the control headquarters server for the information exchange via the WiMAX BTS. These nodes transmit UDP video bit stream data to the server thus providing coverage of the relief activities. The network topology shown in the Figure 1 has been implemented in NS3 simulator. The parameters selected for the mentioned nodes are described in Table 1.

Table 1: Network Environment

| Protocol | Purpose | Nodes | Details |
|---|---|---|---|
| IEEE 802.3 | For backhaul and for relief center local nodes | 10 x Nodes | 5 Mbps nodes. With NS3 CSMA channel between them. The delay of the channel is set to 2 ms. MTU 1400 |
| IEEE 802.11n Wi-Fi | Wi-Fi nodes adjacent to the relief center | 20 x Nodes 1 x AP | WIFI, 80211n, 5GHz |
| IEEE 802.16 WiMAX | For field relief teams | Up to 30 Nodes Modulation | OFDM with 16-QAM, 5GHz |

## 3. Simulation and Results

### 3.1 Experimental Setup

The experimental setup consists of Intel(R) Core-i5 [14], based PC running Microsoft Windows 7 Ultimate Service Pack[15], with 4 GB of system RAM. NS3 needs Linux based platform to run. Ubuntu 10.04 LTS [16], is selected to host NS3 simulator. Ubuntu is installed in VMware Workstation[17]. NS3 does not have a graphical IDE (integrated development environment), although it comes with command line based waf compiler and builder. Linux based ECLIPSE JUNO IDE [18] is used to import the project for development/experimentation. While working on NS3, often, there is need for a light weight IDE for testing small C++ codes. CODEBLOCKS C++ IDE [19], has been used to individually test the segments of C++ code. Table III shows the summary of simulation components.

Table 2: Simulation Setup

| CPU | Intel(R) Core-i5 |
|---|---|
| System RAM | 4GB |
| Host Operating System | Windows 7 |
| NS3 Host Operating System | Ubuntu 10.04 LTS installed on VMware Workstation 7 |
| Simulator | NS3.18 |
| IDE | Eclipse Juno IDE |
| Compiler & Builder | waf |
| HEVC Reference Software | Version HM-9.2 |

The reference software version HM-9.2 has used from the HEVC official online repository. The project is built in Microsoft Visual Studio 2010[20]. For the transmission of HEVC encoded video sequences in NS3, sample sequences have been used from the HEVC official website [21]. These sequences in raw .yuv format are coded using HEVC reference software. The executable encoder and decoder are used with a configuration file to set the desired parameters for the output bit stream. These bit streams thus obtained are used in NS3 simulation.

Table 3: Reference Encoder Settings

| Frame Rate | 24 |
|---|---|
| Frame Size | 832x480, 1920x720, 1280x720, 2650x1600 |
| No of Frames | 10 |
| GoP Size | 4 |
| No of B frames | 3 |
| QP Values used | 22,27,32,37 |

The simulation setup for HEVC video transmission in the NS3 simulator needs modification to the existing NS3 modules. In NS3 packets are sent using dummy data for the simulation. Transmitting coded video data over the simulated topology required that either new application module be written in NS3 or some existing application incorporating dummy data to be appropriately rewritten for serving the purpose. The UDP client server application module of NS3 has been modified for use in the research. The code for UDP client and UDP server application is modified to include file reading and writing in the binary format to read/write HEVC coded bit stream. The bit stream is divided in to suitable length segments which are added as the packet data to be sent by the UDP application. The packets are then sent for transmission over the simulated topology. At the other end of the simulated network the packets are received by the UDP server application. The packet headers are removed and the video coded payload of the packets is retrieved. The video bit stream is then gathered packet wise and reconstructed. The reconstructed *.bin* file is passed through the decoder to ascertain the video quality after being subjected to network errors.

### 3.2 Selection of Network Test Parameters

The disaster area network is characterized by a variety and number of connection and device types. Error and mobility models of NS3 have been used in the simulation. NS3 simulator has elaborate model for inducing error at different nodes in the network. The error induced is used to corrupt the packet contents. These include rate error and burst error models. The rate error model induces error according to the specified rate. The burst error model randomly induces the error in the packets depending upon burst rate and size. In the first simulation, both these error have been applied to the topologies under test. The results with and without the application of the error model have been analyzed.

In the second simulation the number of subscriber stations is varied to observe the effect on delay and jitter performance of the simulated network. The number of nodes was varied by increasing the instances of UDP clients connecting to the central office server application.

NS3 simulator incorporates the mobility model for the Wi-Fi and WiMAX based networks. This model has been used in network simulation scripts. This model sets bounds on the location of nodes inside an imaginary box. The nodes are randomly assigned positions in the bounded box. Then the nodes are given different random speeds. This models the mobility of the real nodes. Random walk mobility model has been used for Wi-Fi nodes in the first simulation. For the second and third simulation constant velocity mobility model has been used. Since the WiMAX nodes support mobility. The built-in constant velocity mobility model of NS3 was used to evaluate the network performance at different speeds while transmitting HEVC content.

Four test video sequences with the same duration (10 frames) have been selected for the analysis. The resolution of the sequences is, Basketball Drill – 832x480, Basketball – 1920x720, Johnny – 1280x720, and Traffic – 2650x1600 respectively. These recommended test video sequences have been used from the official repository of the HEVC. Different resolution and size videos with different motion levels are selected for the experiment. These included 2 basketball sequences with increased motion content, whereas Johnny sequence has lesser motion content. The traffic sequence contained minute details and multiple moving objects.

### 3.3 Results

The first simulation starts with the error performance of transmitted HEVC bit streams. The video sequences presented above are subjected to the simulated network topologies with error induced in the transmission path. NS3 error models of rate and burst error with six values (0.0001, 0.0005, 0.001, 0.005, 0.01 and 0.05 percent), are applied. The results show a little deterioration up to a certain threshold error rate (0.001% in case of rate error model), when the rate error model is used.

The results for the burst error show lesser deterioration for the same percent error rate compared to the rate error model. The burst errors were also more pronounced once they were applied beyond a 0.01 percent error. Figure 2 shows the screen shot results for the Traffic sequence with different error rates and rate error model applied. It can be observed that up to 0.001percent error (Figure 2 (b)) the error effect is acceptable, whereas for 0.01 percent error (Figure 2(c)) the results deteriorate and become unacceptable. However this deterioration depends on the type of frame (I or B type) being corrupted by the error. For I type frames the effect is more pronounced.

The Peak Signal to Noise Ratio (PSNR) and bitrate for the the Basketball drill is shown in Figures 3 and 4. The results in Figure 8 show an overall increase in PSNR after transmission. The initial changes do not affect the received video quality up to 0.001 percent error. However for the error rates greater than 0.005% the PSNR deviates greatly from the original value and result in video deterioration. The bitrate shows no change for the initial error values up to 0.001%. After 0.005 percent error the change in bitrate and deterioration in video quality is almost linear with the percent error.

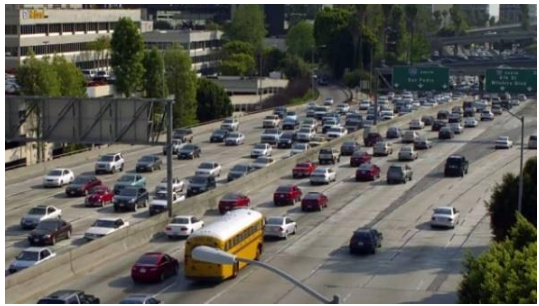
(a)

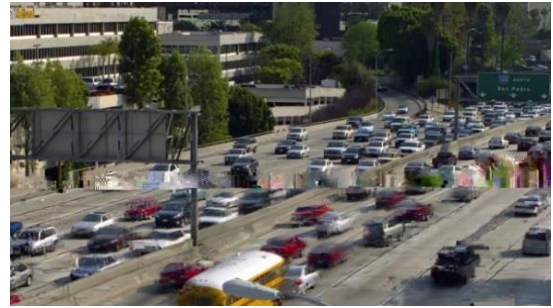
(b)

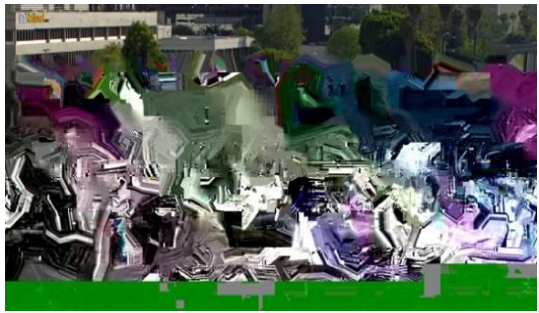
(c)

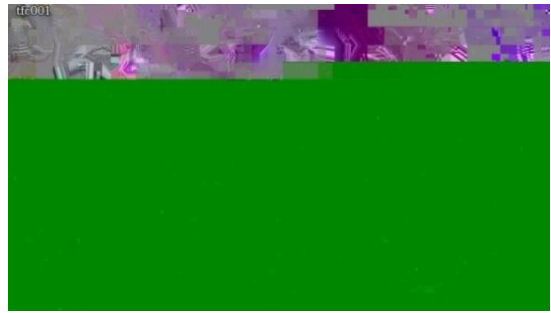
(d)

Fig. 2  Traffic 2560x1600with 0.0001(a), 0.001(b), 0.01(c) and 0.1(d) % Rate Error respectively

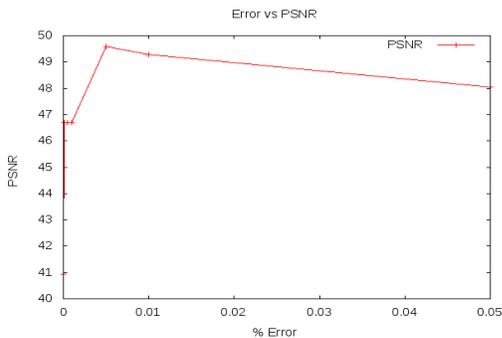

Fig. 3  Percent network rate error Vs YPSNR (20 Nodes)

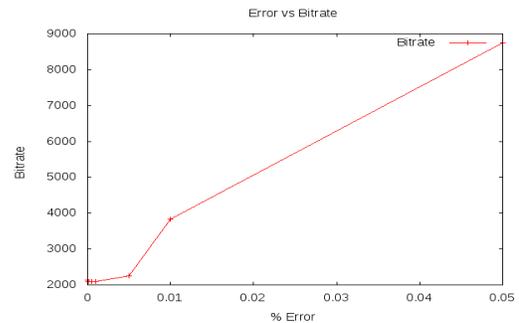

Fig. 4  Percent network rate error Vs Bitrate (20 Nodes)

The second set of simulation tested the performance of the video transfer in terms of delay and jitter of the underlying network. Simultaneous instances of UDP clients transferring the bit streams in same time slots are run. Figure 5 show the number of nodes verses throughput plot for 30 node simulation instance. As can be seen, the increase in number of transmitting nodes causes a gradual decrease in the throughput which finally drops adversely. Here 30 nodes consisting of UDP clients are transmitting to UDP server at the central office.

For streaming applications the general rule is have delay less than 12ms and jitter less than 5ms. Figures 6 and 7, show that the delay and jitter of the network remain within normal bounds until the threshold of 15 transmitting nodes. After that there is sharp increase in the delay and jitter values. It is worth mentioning that for streaming video applications a delay of greater than 12ms affects the video quality. The video streaming application discards the frame after a certain maximum delay. Depending upon the type of frames (for I type the effect is more pronounced), it may have adverse effect on the video quality.

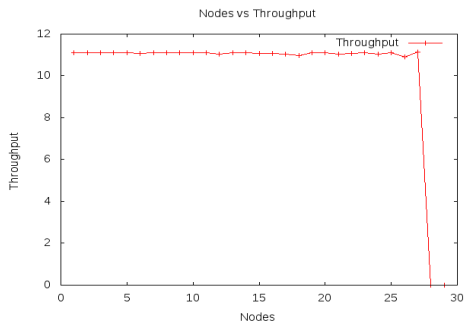

Fig. 5  Nodes Vs Throughput plot for 30 nodes

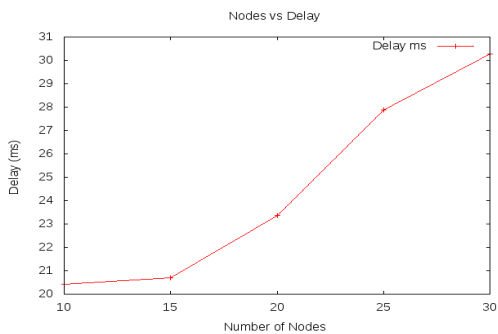

Fig. 6  Number of Nodes Vs Delay for Basketball Drill Sequence

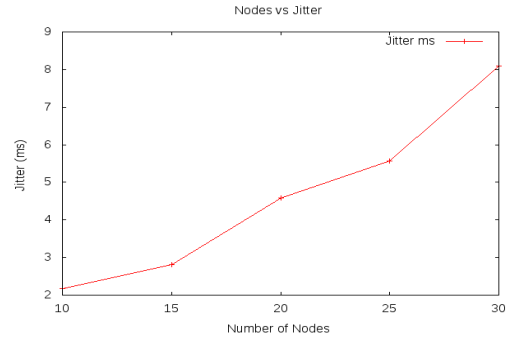

Fig. 7  Number of Nodes Vs Jitter plot for Basketball Drill Sequence

The third set of simulation consisted of evaluation of the mobility performance of the video sequences at different speeds. The simulation consists of mobile WiMAX nodes for which the NS3 constant velocity model is used. The effect of speed on the delay and jitter for 20 WiMAX nodes simulation is shown in Figures 8 and 9. It can be seen that there is negligible effect of speed till 80m/s. However after that there is some fluctuation both in jitter and delay. Although this plot show slight change in both delay and jitter after the increase beyond 80m/s, the resulting packet drops are more pronounced at higher speeds. This shows that despite having the delay and jitter within the acceptable bounds, the packet loss can result at higher speeds. Such packet drops brings visible deterioration in the quality of received video.  Figure 10 show the packet drop statistic at speed of 100m/s. The constant velocity mobility model results in packet drops for the simultaneous packet transfers at speeds beyond 80m/s.

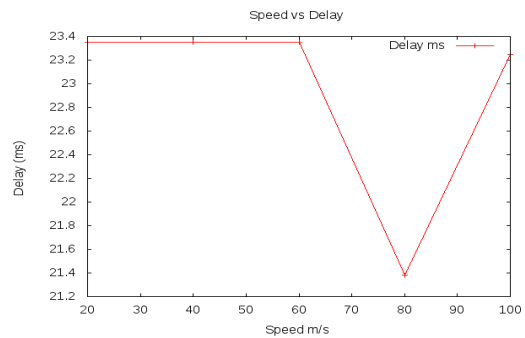

Fig. 8  Number of Nodes Vs Jitter for Basketball Drill Sequence

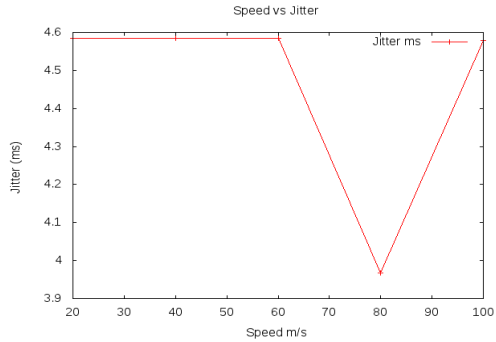

Fig. 9  Speed Vs Jitter for 20 nodes

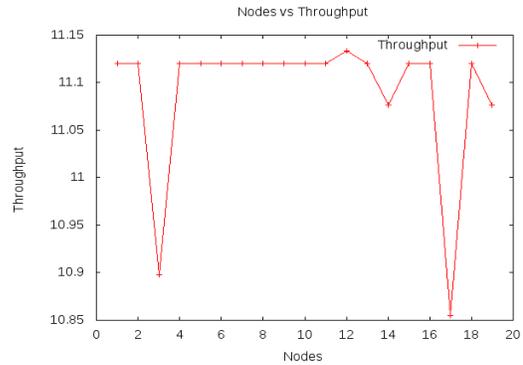

Fig. 11  Number of Nodes Vs Throughput at 100m/s

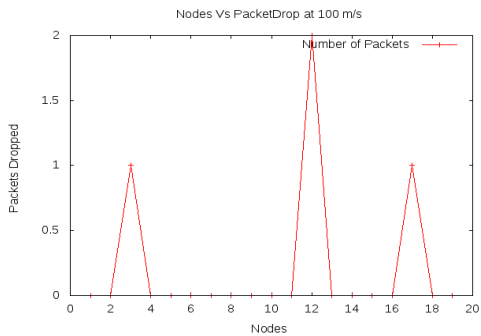

Fig. 10  Nodes-wise Packet Drop at 100m/s

The number of nodes verses throughput plot at speed of 100m/s is shown in Figure 11. As it can be seen the throughput fluctuates for the different flows when tested for 18 simultaneous UDP client nodes.

## 4. Conclusion

In this paper, a disaster area network model for relief activities in calamity stricken locality has been proposed. The performance evaluation of HEVC encoded video content when transmitted over the network has been carried out. The network environment has been simulated using the NS3 simulator and error models from the simulator are used for simulating the effects encountered while transmitting on real networks. It is shown that the HEVC bit streams provide tolerance to network error up to 0.001% of error beyond which the observed quality deteriorate remarkably. The delay and jitter performance of HEVC content transmission stays stable for up to 15 nodes after which there is sharp increase in their values. The performance of HEVC transmission at different speeds shows that beyond 80m/s there is packet drop and the quality deteriorates.

The future work may involve the carrying out the evaluation of real networks. The error performance thresholds for different type of error may further be investigated and error resilience schemes for HEVC in line with previous H264 standard may be developed.  The streaming HEVC video may be used for the future evaluation of the performance of HEVC in similar scenarios.

**Saeed ur Rehman** has received his Bachelors in Electrical Engineering from College of Electrical and Mechanical Engineering, NUST in 2001. He has been working in various engineering setups including automotive and IT related organizations. Presently he is undergoing his MS in Electrical Engineering from UET Taxila, Pakistan. He is a professional member of Pakistan Engineering Council (PEC).

**Dr. Gulistan Raja** is working as Professor in EED, UET, Taxila. He received M.S. Engg., degree from Osaka University, Japan & Ph.D. degree in Electrical Engg., from UET, Taxila. He received Best Teacher Award for year 2009 by Higher Education Commission, Islamabad. He has more than 35 publications in international conferences and journal.